\documentclass[12pt]{article}
\usepackage[english]{babel}
\usepackage{pgf, amsmath, amsfonts, amsthm,amsbsy, mathrsfs, latexsym, multicol,amssymb}
\usepackage{srcltx, hyperref, stackrel, caption}
  \usepackage[rightcaption]{sidecap}
 
\usepackage[margin=3cm]{geometry}

\usepackage{braket}
\usepackage{color}
\usepackage{comment}
\usepackage{graphicx}

\newcommand{\epsi}{\varepsilon}
\newcommand{\E}{{\mathrm{e}}}
\newcommand{\I}{\mathrm{i}}
 \newcommand{\R}{ \mathbb{R} }
\newcommand{\C}{ \mathbb{C} }
\newcommand{\N}{ \mathbb{N} }

\newcommand{\D}{\mathrm{d}}

\newcommand{\tr}{{\mathrm{tr}}}
\newcommand{\trpuv}{\tau}
\newcommand{\Hi}{{\mathfrak{H}}}
\newcommand{\bm}{\begin{pmatrix}}
\newcommand{\Em}{\end{pmatrix}}

\newcommand{\rom}{\renewcommand{\labelenumi}{{\rm(\roman{enumi})}}}

\newcommand{\alp}{\renewcommand{\labelenumi}{{\rm(\alph{enumi})}}}
\newcommand{\balp}{\begin{enumerate}\alp}
\newcommand{\ealp}{\end{enumerate}}

\newcommand{\brom}{\begin{enumerate}\rom}
\newcommand{\erom}{\end{enumerate}}

 \numberwithin{equation}{section}
 
\newcommand{\lminus}{%
  \mathrel{\vbox{\offinterlineskip\ialign{%
    \hfil##\hfil\cr
    $\scriptscriptstyle\Lambda$\cr
    \noalign{\kern-0.2ex}
    $-$\cr
}}}}

\newcommand{\Mminus}{%
  \mathrel{\vbox{\offinterlineskip\ialign{%
    \hfil##\hfil\cr
    $\scriptscriptstyle{M}$\cr
    \noalign{\kern-0.2ex}
    $-$\cr
}}}}

\newcommand{\dimfib}{s}

 \newtheorem{theorem}{Theorem}[section]
  \newtheorem{definition}{Definition}[section]
\newtheorem*{theorem*}{Theorem}

\newtheorem*{lemma*}{Lemma}

\newtheorem{proposition}{Proposition}[section]

\title{Justifying Kubo's formula for gapped   systems at zero temperature: a brief review and some new results}
\author{ Joscha Henheik and Stefan Teufel\thanks{
Fachbereich Mathematik, Eberhard-Karls-Universit\"at\newline
\textcolor{white}{a} \hspace{.7em} Auf der Morgenstelle 10, 72076 T\"ubingen, Germany\newline
\textcolor{white}{a} \hspace{.7em} E-mail:   stefan.teufel@uni-tuebingen.de}}
 
\begin{document}
\maketitle

\begin{abstract}
We first review the problem of a rigorous justification of  Kubo's formula for  transport coefficients in gapped extended Hamiltonian quantum systems at zero temperature. In particular, the theoretical understanding of the quantum Hall effect rests on the validity of Kubo's formula for such systems, a  connection that we review briefly as well. We then highlight an approach to linear response theory based on non-equilibrium almost-stationary states (NEASS) and on a corresponding adiabatic theorem for such systems that was  recently proposed and worked out by one of us in \cite{Teu17} for interacting fermionic systems on finite lattices. 
In the second part of our paper we show how to lift the results of \cite{Teu17} to infinite systems by taking a thermodynamic limit.
\medskip

\noindent \textbf{Keywords.} Linear response theory, Kubo formula, adiabatic theorem, non-equilibrium stationary state.

\medskip

\noindent \textbf{AMS 
Mathematics 
Subject 
Classification (2010).} 81Q15; 
81Q20; 81V70.
\end{abstract}

 \section{Introduction}
 
 In this article we discuss the problem of ``proving Kubo's formula'' for  gapped extended quantum systems at zero temperature, with transport theory in (topological) insulators as the main application in mind. Note that
in this context various expressions are referred to as ``Kubo's formula'', namely the general Kubo formula (KF) for the  response coefficients of arbitrary observables on the one hand, and the double commutator formula (DCF) for the current response on the other.
Thus also different mathematical problems have been subsumed under the label ``proving Kubo's formula''.
As to be detailed below, much attention has been given to the problems of showing that KF implies DCF for the   conductance or conductivity and  to showing that DCF implies (fractional) quantisation of Hall conductance resp.\ conductivity  (see e.g.\ the recent review \cite{Avron} and references therein). Less work has been directed towards a rigorous justification of KF for such systems starting from first principles. As the present work will be mainly concerned with this second problem, let us briefly recall the main challenges such a justification faces.

In the context of Hamiltonian quantum systems,  the linear response formalism for static perturbations answers  the following  question:
How does a system described by a Hamiltonian $ H_0$ that is initially in an equilibrium state $  \rho_0$ respond to a small static perturbation~$ \epsi V$? 
 Or somewhat  more precisely:  What is the change 
\[ 
\tr (\rho_\epsi A) - \tr(\rho_0 A) = \epsi\,\sigma_A + o(\epsi)
\]
of the expectation value  of an observable  $ A$ caused by the perturbation $ \epsi V$ at leading order in its strength $| \epsi|\ll 1$? 
Here $ \rho_\epsi$ denotes the state of the system after the perturbation has been turned on adiabatically  and $  \sigma_A$ is  called the linear response coefficient for~$ A$. The answer clearly hinges on the problem of determining $\rho_\epsi$. While in some situations one expects that $\rho_\epsi$ remains an equilibrium state also for the perturbed Hamiltonian $H_\epsi = H_0 + \epsi V$, 
  Kubo \cite{Kubo} developed linear response theory for situations, where the system is driven into a {\em non-equilibrium state} $\rho_\epsi$.  
  As   emphasized  by Simon  in ``Fifteen problems in mathematical physics'' \cite{S}, 
  the latter situation typically occurs in applications to transport theory and this deviation from equilibrium makes the justification of Kubo's formula   a difficult mathematical problem that is still not solved in satisfactory generality. Also in this work we will not even discuss the general problem but instead focus 
on the rather special situation of  particle transport in gapped Hamiltonian systems (i.e.\ insulators) at zero temperature. In this case  all currents are dissipation-less (direct currents vanish, while Hall currents are geometric and thus dissipation-less) and, as we will argue, a rigorous justification of Kubo's formula  purely on the level of Hamiltonian dynamics without involving any form of dissipation is possible.

The linear response formalism (which we briefly  review in Section~\ref{section2}) rests on the assumption that a {\em small} perturbation ($|\epsi|\ll 1$) that is adiabatically switched on alters the initial equilibrium state of a system only a {\em little} ($\rho_\epsi\approx \rho_0$). 
In a nutshell, the problem of proving Kubo's formula will thus be to prove that a system initially in an equilibrium state $\rho_0$ is adiabatically driven by a small perturbation  into a non-equilibrium state $\rho_\epsi$ close to $\rho_0$.
This problem goes beyond standard   perturbation theory, since  the small perturbation acts over a very long (macroscopic)  time, and thus this assumed small change of the state  is {\em not} a trivial consequence of the smallness of the perturbation; instead, proving this assumption  requires an adiabatic type theorem. 
However,  even if we work at zero temperature and assume that  $\rho_0$ is the gapped ground state of~$H_0$, the problem goes also beyond standard adiabatic theory. Indeed, the standard adiabatic theorem is only of rather limited use here  for three reasons. First, it is only applicable as long as the perturbation does not close the spectral gap; then it asserts that  $\rho_\epsi$ equals  the gapped ground state   of  $H_\epsi = H_0 +\epsi V$ and thus remains an equilibrium state.
But, as emphasized before, in transport theory this assumption is often not satisfied and $\rho_\epsi$ is expected to be a non-equilibrium state;
for example, the band gap in a typical insulator is of order 10eV, while  a macroscopic sample of such a material stays insulating for applied voltages that are larger by many orders of magnitude. 
Secondly, even if one assumes that the spectral gap above the ground state remains open, 
the usual adiabatic theorem is not directly applicable for two reasons: its standard version  estimates the difference between $\rho_\epsi$ and the ground state of $H_\epsi$ in the operator norm; in order to obtain the required estimates with respect to local trace norms, additional and potentially non-trivial   propagation estimates need to be established. Finally,  for extended interacting systems the approximation error in the adiabatic theorem deteriorates when the system size grows and it can not be applied for macroscopic systems. As a consequence, proving Kubo's formula even in the simple case of gapped Hamiltonian systems at zero temperature has been an open problem of mathematical physics for quite some time.

Recently Bachmann et al.\ \cite{BDF} proved an adiabatic theorem for extended interacting lattice systems with error estimates for local traces that are uniform in the system size, thereby solving the second   and third problem. In \cite{Teu17} one of us solved also the first problem by proving a version of the adiabatic theorem that remains valid   even when the perturbation closes the spectral gap. In a nutshell, the idea in \cite{Teu17} is that perturbations by slowly varying but not small potentials (modelling small fields acting on large regions of space) close the spectral gap, but leave intact a local gap structure, thereby driving the system into a non-equilibrium almost-stationary  state (NEASS). The  adiabatic theorem in \cite{Teu17} states that if such a perturbation is switched on adiabatically, then the state of the system evolves into a uniquely determined NEASS associated with  the perturbed Hamiltonian that has an explicit asymptotic expansion in powers of $\epsi$; \cite{Teu17} also applies to interacting extended lattice systems and provides error estimates that are uniform in the system size.
This NEASS approach was motivated by and is based on ideas which we called space-adiabatic perturbation theory almost 20 years ago, see for example \cite{PST1,PST2,T}, and allows not only to prove validity of Kubo's formula, but also to evaluate it in a straightforward way. The latter point is illustrated in \cite{MMPT}, where we compute the spin-Hall conductivity in topological insulators.

The rest of this note is structured as follows.  Section~\ref{section2}  first recalls the formal derivation of Kubo's formula in the context of Hamiltonian quantum systems and   highlights the different mathematical problems arising from it.  In particular, we try to clarify the different meanings that the phrase ``proving Kubo's formula'' acquired in the past.  We then try to provide a concise and structured overview of the mathematical literature in this area.

In Section~\ref{section3},    we discuss  the extension of the results in \cite{Teu17} to infinite systems.
We show that NEASSs exist as states on the quasi-local algebra of the infinite systems and are automorphically equivalent to the ground state of the unperturbed Hamiltonian, then state a version of a corresponding adiabatic theorem, and finally  obtain from there a rigorous justification of  an infinite volume version of Kubo's formula. Proofs and generalizations of the results presented in Section~\ref{section3} will be given elsewhere \cite{HT}.\\

\noindent{\small{\bf Acknowledgements:} 
I (S.T.) would like to thank   
Yosi Avron,
Sven Bachmann,
Horia Cornean,
Giuseppe De Nittis,
Wojciech De Roeck,
Alexander Elgart,
Martin Fraas,
J\"urg Fr\"ohlich,
Vojkan Jaksic,
Max Lein,
Giovanna Marcelli,
Domenico Monaco,
Gianluca Panati,
Marcello Porta,
and
Marcel Schaub
for sharing their insights and ideas about this complex topic with me and/or  for critically commenting on some of my own ideas.
}
  \section{Linear response: heuristics, problems, and   results}\label{section2}

In this section we first recall the standard derivation of the general Kubo  formula for Hamiltonian quantum systems and static perturbations.  
We then highlight the steps in the derivation that require a more careful justification. In Subsections~\ref{sectionA}--\ref{sectionC} we discuss some  existing (mostly) mathematical literature addressing the different aspects of the problem. 

Consider a quantum system described by the self-adjoint  
Hamiltonian $H_0$ on some Hilbert space $\mathcal{H}$  that is bounded from below and subject to a   perturbation $\epsi V$ with $|\epsi|\ll 1$ such that also $H_\epsi = H_0 + \epsi V$ is self-adjoint. The simplest example to keep in mind would be a single atom perturbed by a small external field, while   relevant to transport theory are for example Hamiltonians describing  
fermions on a lattice with short range interactions subject to a perturbation  by a small external field. 
Assume that initially, i.e.\ before the perturbation is applied, the system is in an equilibrium state $\rho_0$, i.e.\
$\rho_0 \sim \E^{-\beta H_0}$ if the temperature $T= (k_{\rm B}\beta)^{-1}$ is positive, or $\rho_0$ equal to the ground state of $H_0$ if the temperature is zero.  
The objective of linear response theory   is to determine the change of expectation values of observables  $A$ linear in the strength of the applied perturbation, 
\begin{equation}\label{linres1}
\langle  A\rangle_{\rho_\epsi}  - \langle  A\rangle_{\rho_0}:=
\tr (\rho_\epsi A) - \tr(\rho_0 A) =    \epsi  \, \sigma_A + o(\epsi)\,.
\end{equation}
Here $\rho_\epsi$ is the state of the system after the perturbation has been turned on and $\sigma_A$ is the linear response coefficient for the observable $A$ with respect to the perturbation $V$. The   question is now: What is the state $\rho_\epsi$?   To answer this question, one gets back to first principles and models the time-dependent switching of the perturbation by solving the corresponding time-dependent Schr\"odinger equation.
Assume that the switching occurs during the time interval $[-1,0]$ and the Hamiltonian at time $t$ is 
\[
H_\epsi(t) = H_0 + f(t) \,\epsi V
\]
with a smooth switching function $f: \R \to [0,1]$ such that $f(t) = 0$ for $t\leq -1$ and $f(t) =1$ for $t\geq 0$.
Then the state of the system at time $t$ is given by the solution $\rho_\epsi(t)$ to the time-dependent Schr\"odinger equation (we choose units where $\hbar=1$)
\begin{equation}\label{Schroe1}
\I \tfrac{\D}{\D t} \rho_\epsi(t) =  [ H_\epsi(t), \rho_\epsi(t)] \qquad \mbox{with }   \rho_\epsi(t) = \rho_0 \mbox{ for all } t\leq -1\,.
\end{equation}
Hence, if one measures the observable $A$ at time $t\geq 0$ after the perturbation is fully switched on, one should use the state $\rho_\epsi = \rho_\epsi(t)$ in \eqref{linres1}. Standard time-dependent perturbation theory yields 
\[
\rho_\epsi(t) \;=\; \rho_0   - \epsi \, \I  \int_{-\infty}^t f(  s) \, \E^{ \I H_0  (s-t)}\, [V, \rho_0]  \, \E^{-\I H_0  (s-t)}\D s + R(\epsi ,f, t)\,,
\]
with a remainder term $R(\epsi ,f, t)$ that is $o(\epsi)$ in a   sense to be discussed, and thus
\begin{eqnarray*}
\langle  A\rangle_{\rho_\epsi(t)}  - \langle  A\rangle_{\rho_0} 
 &=& - \epsi \, \I  \int\limits_{-\infty}^t  f(s)  \, \tr \left( \E^{ \I H_0  (s-t)}\, [V,\rho_0] \, \E^{-\I H_0  (s-t)}  \,A\right)\D s\;   +\;   \tr(R(\epsi ,f, t) A) \,.
\end{eqnarray*}
 As the perturbation acts only during a finite time-interval, one might  expect\footnote{As to be discussed below, already  the proof of  this step can be technically quite demanding for several reasons, one being the  control of the trace norm of $R(\epsi,f,t)$ instead of the operator norm.} 
 that $\lim_{\epsi\to 0} \epsi^{-1} \tr(R(\epsi ,f, t)A) = 0$ and thus that \eqref{linres1} indeed holds with
\begin{equation}\label{Kubo1}
\sigma_A \;= \;\sigma_A^{f}(t) 
\;=\;
 -  \I  \int\limits_{-\infty}^t  f(s)  \, \tr \left( \E^{ \I H_0  (s-t)}\, [V,\rho_0] \, \E^{-\I H_0  (s-t)}  \,A\right)\D s \,.
\end{equation}
However, the response coefficient $\sigma_A^f$ defined in this way would generically depend on the switching function $f$   and also on the time $t$, even for $t\geq 0$. In particular, one could  not hope for a simple universal formula for it. But one expects in many relevant situations that response coefficients are independent of experimental details like the exact way of how to turn on an external perturbation or the   time at which the measurement takes place after the perturbation has been turned on. Thus, from a practical view-point,  \eqref{Kubo1} is clearly an unsatisfactory definition.

One solution to this problem is provided by taking  an adiabatic limit: Since the time-scale on which the perturbation is applied is typically long compared to the internal time-scales of the quantum system, one  considers \eqref{Schroe1} in the adiabatic limit, i.e.\ one considers the limit of   slow switching. Introducing the adiabatic parameter $0<\eta\ll 1$, the adiabatic Schr\"odinger equation 
\begin{equation}\label{Schroe2}
\I \tfrac{\D}{\D t} \rho_{\epsi,\eta}(t) =  [ H_\epsi(\eta t), \rho_{\epsi,\eta}(t)] \qquad \mbox{with }   \rho_{\epsi,\eta}(t) = \rho_0 \mbox{ for all } t\leq -1/\eta \,,
\end{equation}
describes the same switching process  but stretched to the longer  time-interval $[-1/\eta,0]$.
The hope is now that in the adiabatic regime $\eta\ll 1$   the response coefficient  \begin{equation}\label{Kubo2}
 \;\sigma_A^{f,\eta }(t) 
  \;:=\; - \I  \int\limits_{-\infty}^t  f(\eta s)  \, \tr \left( \E^{ \I H_0  (s-t)}\, [V,\rho_0] \, \E^{-\I H_0  (s-t)}  \,A\right)\D s    
\end{equation}
becomes independent of $f$,   $\eta$, and also of $t$, whenever $t\geq 0$. Replacing moreover the generic switch function $f$ by an exponential function and evaluating at $t=0$,  the integral in \eqref{Kubo2} becomes the Laplace transform of the Heisenberg time-evolution and thus the resolvent of its generator, the   Liouvillian $V\mapsto \mathcal{L}_{H_0} (V)  :=  [H_0,V]$. 
One thereby arrives at  the general Kubo  formula (KF)  for the linear response coefficient of an observable $A$,\footnote{It should be noted that  in general the limit $\eta\to0$ in \eqref{Kubo3} need not  exist and sometimes $\eta>0$ is taken as an empirical parameter that controls the strength of dissipation in the system. However, in our setting of gapped systems at zero temperature the   limit $\eta\to0$ in \eqref{Kubo3}
is expected and can be shown to exist in many models.
}
\begin{eqnarray}\label{Kubo3}
\sigma_A^{\rm Kubo}  &:=  &-\I\, \lim_{\eta\to 0^+}     \int\limits_{-\infty}^0  \E^{\eta s}  \, \tr \left( \E^{ \I H_0  s}\, [V,\rho_0] \, \E^{-\I H_0  s}  \,A\right)\D s   \nonumber\\
 &=& 
  \lim_{\eta\to 0^+}  \tr \left(   (  \mathcal{L}_{H_0}-\I  \eta)^{-1} ([V,\rho_0]) 
A\right)\,.
\end{eqnarray}
This heuristic derivation   immediately leads to two questions:
\begin{enumerate}
\item[(A)] Under which assumptions on the model (Hamiltonian $H_0$, perturbation $V$, observable $A$, initial state $\rho_0$) does the right hand side of \eqref{Kubo3} lead to a well defined number $\sigma_A^{\rm Kubo}$ and how can it be evaluated more explicitly for   current observables to obtain, for example,  the double-commutator formula (DCF)  sometimes called Kubo-Streda formula? \item[(B)] Assuming that \eqref{Kubo3} leads to a well defined number $\sigma_A^{\rm Kubo}$, under which additional assumptions on the model (Hamiltonian $H_0$, perturbation $V$, observable $A$) is $\sigma_A^{\rm Kubo}$   a universal linear response coefficient? I.e., when is it true that for all smooth switching functions $f$ and all times $t\geq 0$ one has 
\begin{equation}\label{B1}
\langle  A\rangle_{\rho_{\epsi,\eta}(t)}  - \langle  A\rangle_{\rho_0}    = \epsi \cdot \sigma_A^{\rm Kubo} + R(\epsi,\eta, f, t) 
\end{equation}
with
\begin{equation}\label{limitremainder}
\lim_{\epsi   \to 0} \;\sup_{\eta \in I_\epsi} \;\frac{R(\epsi,\eta,f ,t)}{\epsi} = 0
\end{equation}
for some interval $I_\epsi \subset (0,\infty)$ of admissible time-scales $\eta$.
As we will argue,   for dissipation-less currents, because of tunnelling, it is not   expected  that the supremum in  \eqref{limitremainder} can be replaced by a limit $\eta\to 0$ in general, cf.\ also Theorem~\ref{response} and the remark afterwards.  Moreover, for extended interacting systems one also needs to show that \eqref{limitremainder} holds uniformly in the number $N$ of particles, i.e.\ that
\begin{equation}\label{limitremainderN}
\lim_{\epsi   \to 0} \;\sup_{\eta \in I_\epsi}\; \sup_{N\in \N} \;\frac{R(\epsi,\eta,f ,N,t)}{\epsi} = 0
\end{equation}
as otherwise the estimate  may deteriorate and become worthless in the thermodynamic limit.
\end{enumerate}
From now on we only discuss the situation of gapped Hamiltonians at zero temperature.
Still, the mathematical difficulty of both problems (A) and (B) depends very much on various details of the specific model under consideration:
\brom
\item Does $H_0$ describe particles  on a lattice or in the continuum? 
Lattice Hamiltonians are typically bounded self-adjoint operators, while continuum Schr\"odinger operators are unbounded. The same distinction then holds for the associated current observables. 
\item Are the particles interacting or non-interacting? For non-interacting particles one can consider the one-body Hamiltonian on an infinite domain;  then controlling estimates uniformly in the number of particles is no issue. Interacting systems need to be first analyzed on finite domains and the thermodynamic limit becomes a nontrivial step. 
\item  Is $H_0$ assumed to have a spectral gap at the Fermi energy resp.\ above the ground state, or only a mobility gap? The first situation occurs typically if  $H_0$ is (a small perturbation of) a periodic non-interacting Hamiltonian, while the second situation is expected to occur for generic random Hamiltonians.
\item Does the perturbation $V$ close the spectral resp.\ mobility gap? Perturbing by a constant electric field $E$, i.e.\ by a linear potential $V(x) = x\cdot E$ will typically close all spectral or mobility gaps of $H_0$,  no matter how small $\epsi$ is. On the other hand, for any bounded perturbation $V$ the gap remains open for $\epsi$ small enough.
\item Is the observable under consideration local or extensive?   In the latter case some notion of trace per unit volume needs to be established in order to handle infinite domains or the thermodynamic limit.  
\erom
All aspects in the above list have been addressed in some form or another for  problem~(A). Although (A)  is not the main focus of this note, we briefly sketch the problem in  Subsection~\ref{sectionA} and mention some literature. 
Problem (B) has attracted much less attention. In Subsection~\ref{sectionB} we first discuss the problem  on a heuristic level and then mention the few  existing mathematical results. Subsection~\ref{sectionC} collects references to further mathematical works in the context of linear response for extended quantum systems.

\subsection{Evaluating     Kubo's formula for the current observable}\label{sectionA}

Evaluating Kubo's formula \eqref{Kubo3} for current observables can be    tricky. To see this, first note that on finite domains $\Lambda\subset\R^d$  the total current response always vanishes because of conservation of total charge. This follows also easily by evaluating Kubo's formula:
Let the perturbation $V$ be the potential of a constant electric field  of unit strength  pointing in the $i$th coordinate direction, i.e.\ $V= X_i$ with $X_i$ the $i$th component of the position operator and  the observable $A$  the current in  the $j$th coordinate direction, $A=J_j := \I[H_0, X_j] = \I \mathcal{L}_{H_0}(X_j)$. Then a naive evaluation of 
\eqref{Kubo3} yields
\begin{eqnarray}\label{DK1}
\sigma_{ij}^{\rm Kubo}
   &=&  \I \, \lim_{\eta\to 0}  \tr \left(   (  \mathcal{L}_{H_0}-\I  \eta)^{-1} ([X_i,\rho_0]) \,
  \mathcal{L}_{H_0}(X_j) \right)\nonumber\\
   &=& 
\I \, \lim_{\eta\to 0}  \tr \left(  [X_i,\rho_0]  (  \mathcal{L}_{H_0}+\I   \eta)^{-1} (
   \mathcal{L}_{H_0}(X_j) )\right)\\
  &=& 
\I \,  \tr \left(  [X_i,\rho_0]   X_j \right)\;=\;  - \I \,\tr \left( \rho_0 \left[ X_i , X_j \right]\right)=0\,.\nonumber
\end{eqnarray}
Whenever $X_i$ and $X_j$ are trace class, which is the case  in lattice models on bounded domains, then the above computation is perfectly valid and, as expected, the current response  vanishes. 
In order to see a nontrivial  current response, one thus either   works on an infinite domain, or on a domain with a torus geometry, or considers only local currents. 
In the latter cases, to avoid finite size or boundary effects, one eventually would like to take a thermodynamic limit as well. Thus, in all cases, additional mathematical challenges appear when trying to evaluate   Kubo's formula.

\subsubsection{The double-commutator formula and quantization for non-interacting systems on infinite domains}
For non-interacting fermionic systems one can directly work with the one-body Hamiltonian $H_0$. Then, at zero temperature,   the initial state is  given by the corresponding Fermi projection  $\rho_0 = \chi_{(-\infty, \mu]}(H_0)$. In order to evaluate extensive observables like the current as densities, one needs to establish the notion of a trace per unit volume $\trpuv$ which is cyclic for operators in suitable trace or Hilbert-Schmidt classes. For   random ergodic systems, a corresponding mathematical formalism   has been worked out by Bellissard, van Elst, and Schulz-Baldes \cite{BES} in the discrete case (see also the work of Aizenman and Graf \cite{AG} for a different perspective), by Bouclet, Germinet, Klein, and Schenker  \cite{BGKS} for random Schr\"odinger operators, and by De Nittis and Lein \cite{DL} for a large abstract class of operators including all previous ones.

Even in the discrete case  the position operators are unbounded and, in parti\-cular, not in any  trace per unit volume class.  As a consequence, the naive computation \eqref{DK1} fails to produce the correct result. 
For a correct evaluation of Kubo's formula one observes that, since $\rho_0$ is now a projection,    in the first line of  \eqref{DK1} only 
the off-diagonal part 
\[
X_i^{\rm OD} := \rho_0 X_i (1-\rho_0) +   (1-\rho_0) X_i \rho_0 %= [ [X_i,\rho_0],\rho_0]
\]
contributes, which can be shown to be
 periodic   resp.\ covariant for periodic resp.\ random ergodic  Hamiltonians. And then simple algebra shows that also $X_j$ can be replaced by $X_j^{\rm OD}$. From this one finds as in \eqref{DK1} the celebrated double-commutator formula for the conductivity tensor
\begin{equation}\label{DK2}
\sigma_{ij}^{\rm Kubo} =
- \I\, \trpuv \left( \rho_0 \left[ X_i^{\rm OD} , X_j^{\rm OD} \right]\right) 
=  \I\, \trpuv \left( \rho_0 \left[ [\rho_0,X_i]  , [\rho_0,X_j]\right]\right) .
\end{equation}
A version of this formula in terms of an integral of derivatives of Bloch functions over the Brillouin torus appeared first in the work of   Thouless,  Kohmoto,   Nightingale, and  den Nijs \cite{TKNN}  for periodic Hamiltonians. The authors of \cite{TKNN}  realized that the resulting expression is quantized, leading to the  first understanding of the integer quantum Hall effect in terms of a microscopic model and to a Nobel prize in physics for  David Thouless in 2016. 
The explicit form of \eqref{DK2} in terms of a double commutator  was first given by Avron, Seiler, and Simon \cite{ASS} and  its interpretation as  the Chern number of a complex line bundle over the Brillouin torus by Simon in \cite{SChern}.
One should mention  that at the same time as \cite{TKNN} independently also Streda \cite{Streda} found a   formula for the Hall conductivity from which  he could conclude quantization (see also Subsection~\ref{sectionC}).

However, while expressing the Hall conductivity explicitly in terms of a topological quantity was a major step towards a microscopic understanding of the integer quantum Hall effect, an important ingredient was still missing:
 In order to understand the quantized plateaux appearing in experiments, disorder and the resulting Anderson-localized states appearing in the gap need to be taken into account. On a rigorous level for infinite domains, this was first achieved by Bellissard, van Elst, and Schulz-Baldes in \cite{BES} based on the earlier idea by Bellissard \cite{Be88} to use the framework of non-commutative geometry for extending the work of TKNN to non-periodic systems with mobility gap. More precisely, in \cite{BES} not only a $C^*$-algebraic framework is developed that allows to derive the double-commutator formula from Kubo's formula and to show that it agrees with Streda's formula, but it is also shown that if the Fermi energy lies in a mobility gap, then the double-commutator formula leads to a quantized result for the Hall conductivity    and a zero result for the direct conductivity. To understand why the latter remains exponentially small as a function of  the inverse temperature when dropping the idealization of zero temperature is an important  problem in its own, cf.\ \cite{ABS}.

 Bouclet, Germinet, Klein, and Schenker \cite{BGKS}  set up linear response theory and  derive the double-commutator formula for random Schr\"odinger operators. Here the main challenge was to develop the algebraic-analytic framework for defining the trace per unit volume and associated trace ideals. The authors also derive Kubo's formula starting from time dependent perturbation theory, however, with two caveats: First, \eqref{limitremainder} is shown only for fixed adiabatic parameter $\eta$. Uniformity in $\eta$ for systems with mobility gap is still a completely open problem. Second, the linear time-dependent perturbing potential $\epsi f(t) X_i$ is replaced by a time-dependent vector potential $A(t) = \epsi \int_{-\infty}^t f(s)\D s\, e_i$. While formally the two problems are related by a time-dependent gauge-transformation, translating their results back to the original setting is technically demanding because of subtle domain issues. 
 Moreover, as we will argue below,  understanding linear response in the gauge with linear electric potential also sheds some light on the physics.
 
  Based on the mathematical framework developed in \cite{BGKS}, Klein, Lenoble, and M\"uller  \cite{KLM} also evaluated Kubo's formula for the AC-conductivity and rigorously found Mott's formula for its asymptotic behavior at low frequencies.
Finally, De Nittis and Lein \cite{DL} further generalized the framework of \cite{BGKS} to cover an extremely general class of unbounded operators.

\subsubsection{The double-commutator formula and quantization for interacting systems}

For interacting systems one starts out with a family of Hamiltonians $H_0^\Lambda$ parametrized by   finite domains $\Lambda$. As explained above, if one is interested in a non-trivial current response, $\Lambda$ should be taken as a torus, i.e.\ the cube  $[-L/2,L/2]^d$ is understood with suitable periodic boundary conditions. One fixes the density $\varrho$ by relating the number of particles to the volume $|\Lambda|= L^d$ as $N/|\Lambda| =\varrho$.
The initial equilibrium state at zero temperature is now the ground state $\rho_0$ of $H_0$. In all the mathematical results to be discussed in the following, one assumes that the ground state is separated by a gap from the remainder of the spectrum uniformly in the size of the system. This assumption corresponds to the gap assumption for the one-body Hamiltonian and
no mathematical results analogous to the ones presented below exist for  interacting systems with   a mobility gap above the ground state.

Historically the first work on understanding quantization of charge transport in terms of a microscopic model for interacting particles is Niu and Thouless \cite{NT}. They derive an expression for the transported charge in one cycle  from the adiabatic response of such a system under periodic driving by an external field. This expression has again the interpretation of a Chern number of a line bundle over a two-dimensional torus. One direction on the torus is time (one period), the other is a complex phase characterizing the boundary conditions. Only by averaging over time and over boundary conditions quantization can be concluded. However, assuming that in the thermodynamic limit the value of the transported charge is independent of the chosen boundary condition, also quantization of transported charge   without averaging would follow for large systems.

Shortly after, 
in 1985 Niu, Thouless, and Wu \cite{NTW}, and independently   Avron and Seiler \cite{AS} formulated   similar arguments showing quantization of a suitably averaged Hall conductivity resp.\ conductance  
for interacting electron systems. While Niu et al.\ consider the conductivity and average over a family of boundary conditions parametrized by a two-torus, Avron and Seiler come back to Laughlin's original argument for the conductance and average over a flux torus. In both works the averaging over the corresponding  torus yields again a double-commutator formula for the conductivity resp.\ conductance that has the geometric meaning of a Chern number and implies quantization. Also, in both works  the validity of Kubo's formula \eqref{Kubo3} is taken for granted.

Only 30 years later Hastings and Michalakis \cite{HM} were able to prove rigorously that in the case of Hall conductance the averaging over the flux torus is not needed for large systems.  More precisely, they show that the Hall conductivity  for a gapped Hamiltonian is quantized up to terms that are asymptotically smaller than any inverse power of the size $L$ of the system. Recently, the argument was considerable simplified and generalized by Bachmann, Bols, De Roeck, and Fraas \cite{BBDF}.

A different and more general perspective on the (fractional) quantum Hall effect was developed in a series of works by Fr\"ohlich and collaborators   (see e.g.\ \cite{FK} and the recent review \cite{Fr}). They show that the large-scale properties of  two-dimensional electron gases with vanishing longitudinal resistivity  are governed by  effective Chern-Simons gauge theories.  From the latter fact all the phenomenology of (fractional) quantum Hall systems can be derived. The assumption of 
vanishing longitudinal resistivity (which is equivalent to vanishing longitudinal conductivity in two-dimensional systems) follows from Kubo's formula when assuming a (mobility) gap. Thus proving Kubo's formula seems relevant also to their approach.

Yet another    route to proving quantization of Hall conductivity in interacting lattice systems was taken by 
Giuliani, Mastropietro, and Porta \cite{GMP}. They start from a gapped periodic system of non-interacting  fermions (for which quantization of Hall conductivity is understood since the work of Thouless et al.\ \cite{TKNN}) and use cluster expansion techniques to show that the Hall conductivity does not change when a sufficiently small interaction between the electrons is added to the Hamiltonian. While validity of Kubo's formula  is taken for granted also here, this approach does not assume stability of the spectral gap under small perturbations.

\subsection{Justifying   Kubo's formula for gapped Hamiltonian systems}\label{sectionB}

In this section we address Problem (B) in some more detail. We  start with some heuristics and then briefly describe existing results.

\subsubsection{Justifying   Kubo's formula: Heuristics}

Under what conditions  do we expect Kubo's formula \eqref{Kubo3} to yield a universal response coefficient in the sense that \eqref{B1} holds uniformly as expressed in \eqref{limitremainder} or \eqref{limitremainderN}? 
The idea behind the adiabatic switching procedure is that  for $0<\eta\ll1$ the state $\rho_{\epsi,\eta}(t)$ of the system (as determined by the Schr\"odinger equation \eqref{Schroe2}) follows closely a curve $\Pi_\epsi(t)$ of (almost)-stationary states for  the instantaneous Hamiltonians $H_\epsi(t)= H_0 + f(t)\epsi V$ at all times. If $\Pi_\epsi(t)$ depends only on the instantaneous Hamiltonian $H_\epsi(t)$ and not on $f$ and $\eta$, this would explain why the response of the system after the perturbation $\epsi V$ is fully applied   is   independent of time $t\geq 0$ and of  the details of the switching procedure, i.e.\ of $f$ and $\eta$. Moreover, the derivation of Kubo's formula also rests on the assumption that $\rho_{\epsi,\eta}(t)$ deviates only little from $\rho_0$. Hence, for  linear response theory to work as intended, there should be  non-equilibrium (almost-)stationary states $\Pi_\epsi(t)$ for  $H_\epsi(t)$  (let us call them NEASS in the following)
that are small perturbations of $\rho_0$ such  that $ \rho_{\epsi,\eta}(t)\approx \Pi_\epsi(t)$  in an appropriate sense for $\eta$ sufficiently small.

If the perturbation $\epsi V$ does not close the spectral gap of $H_0$, then the instantaneous ground state of $H_\epsi(t)$ is the natural candidate for $\Pi_\epsi(t)$. Indeed, in this case the adiabatic theorem of quantum mechanics implies  immediately that $\lim_{\eta\to 0} \rho_{\epsi,\eta} (t)= \Pi_\epsi(t)$ in norm. But this means that for $0<\eta\ll 1$ the state $\rho_{\epsi,\eta}(t)$ of the system follows a curve of equilibrium stationary states and ends up in the zero temperature equilibrium state of the perturbed Hamiltonian $H_\epsi=H_\epsi(0)$. While proving \eqref{Kubo3} with \eqref{limitremainderN} is still a highly non-trivial task in this case (as to be discussed below), from a physics perspective this result falls short of justifying linear response in its intended generality. As emphasized in the introduction, linear response theory is designed to provide response coefficients  specifically also in those cases, where the system is driven out of equilibrium.

As we will explain next,  in our setting of gapped Hamiltonians $H_0$  (describing insulating materials)  there is indeed a clear and simple physical picture that suggests the existence of NEASS for $H_\epsi(t)$ even when a perturbation $\epsi V$     being the (possibly unbounded) potential of an external small electric  field   closes the spectral gap of $H_0$. Assume for simplicity that $H_0$ is a periodic one-body operator  in dimension $d=1$ and that the Fermi energy $\mu$ lies in a spectral gap of size $g$. Then in the initial state $\rho_0$ all one-body states with energy smaller than $\mu$ are occupied and it takes at least energy $g$ to excite one electron from the filled bands to an empty band. 
 \begin{SCfigure}[1.15]
   \includegraphics[width=0.5\textwidth]{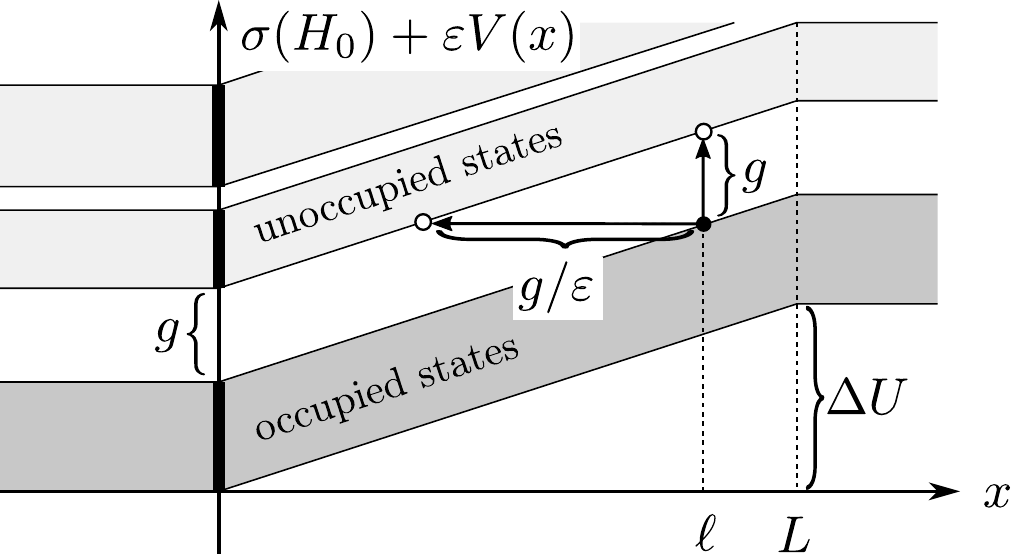}
    \caption{\small Transitions from the occupied bands into the unoccupied bands are  strongly suppressed: To make a transition into an unoccupied state, an electron in the occupied band (black dot) must either overcome the energy gap $g$ (vertical arrow) or tunnel a distance of order $g/\epsi$ (horizontal arrow).      } 
    \label{figure1}
\end{SCfigure}
If one   applies a voltage $\Delta U$ over a distance $L$ modeled by the addition of  a   potential $V_{\Delta U, L} (x) = \frac{\Delta U}{L} \left( x \chi_{[0,L]}(x) + L  \chi_{[L,\infty)}(x)\right) $, 
then for $\Delta U >g$ the perturbed Hamiltonian will no longer have a spectral gap at $\mu$. 
Still,  the local field strength $\epsi:=  \frac{\Delta U}{L}$ in the region $[0,L]$ is  small if $L$ is a macroscopic distance. 
For an  electron  in the Fermi sea that is localized in a microscopic region around the point $\ell\in [0,L]$ the addition of the potential  might result in a substantial shift $\epsi\ell$ in energy, but it still experiences a small force of order $\epsi$. And in order to make a transition into an unoccupied state, it still needs to either overcome the energy gap  of size $g$, or  tunnel a large distance of order $g/\epsi$, see Figure~\ref{figure1}. Thus, as long as $g/\epsi$ is large, the state $\rho_0$ is still an almost-stationary state for the perturbed Hamiltonian $H_\epsi = H_0 + V_{\Delta U, L} $. However,  $\rho_0$ is certainly neither close to the  ground state nor to any other type of  equilibrium state of $H_\epsi$. Note that this heuristic picture still remains valid when local interactions between the electrons are taken into account. 

At this point let us briefly mention the analogy to shape resonances in the case of finite systems. As the simplest example consider the hydrogen atom in a constant electric field, i.e.\ the Stark Hamiltonian. No matter how small  the electric field is, the spectrum of the Stark Hamiltonian is the whole real line. However, when the field is of order $0<|\epsi|\ll 1$,  then  the ground state of the hydrogen Hamiltonian  is close to a shape resonance of the Stark Hamiltonian that is stable for very long times.  
And when one adiabatically turns on a small electric field, one expects the initial ground state of a hydrogen atom to adiabatically evolve into this resonance. Note that adiabatic theorems for resonances have been established   e.g.\ in \cite{AF,EH}, although for technical reasons they do not cover the case of the Stark Hamiltonian.

The situation described above for extended non-interacting periodic systems is very similar: instead of a simple ground state the unperturbed state $\rho_0$ is the spectral projection of $H_0$ onto an infinite dimensional spectral subspace separated by a gap from the rest of the spectrum. And the NEASS described above could be seen as an infinite dimensional resonance for $H_\epsi$. For interacting extended systems, the analogy is even stronger, as $\rho_0$ is indeed the gapped ground state of $H_0$ and the corresponding NEASS could also be called a resonance of $H_\epsi$. 

Thus, in order to prove Kubo's formula also for perturbations that close the spectral gap, one needs to first understand the NEASSs (or ``resonances'') of $H_\epsi$ described above, and then to prove an adiabatic type theorem for NEASSs. This was done in \cite{Teu17}, and in Section~\ref{section3} corresponding results in the thermodynamic limit are discussed.

\subsubsection{Justifying Kubo's formula: Mathematical results}

The first  work concerned with a rigorous justification of Kubo's formula in the sense just described, and that  we know of,  is by
Elgart and Schlein \cite{ES}. They  consider  non-interacting electrons on $\R^2$ described by  the Landau Hamiltonian with smooth and  bounded potential and the Fermi energy in a gap of the Hamiltonian. They derive Kubo's formula for what is often referred to as  conductance in this context by proving an appropriate adiabatic theorem. Note that for computing the conductance, one applies the electric field and measures the current only locally, i.e.\ one  replaces  $X_i$ and $X_j$ in our   discussion of conductivity above  by $G(X_i)$ and $G(X_j)$, where $G:\R\to [0,1]$ is a smooth step function with compactly supported derivative~$G'$. This results in two technical simplifications:  no notion of trace per unit volume is needed, and the perturbation is bounded and therefore the gap remains open for $\epsi$ sufficiently small. Thus in \cite{ES} the standard adiabatic expansion with open gap can be applied and the main technical challenge is to prove the propagation estimates needed for controlling the error in local trace norms (see also \cite{Marcelli}). It is also worthwhile to mention  that in \cite{ES} the adiabatic parameter $\eta$ and the perturbation parameter $\epsi$ are identified, $\eta =\epsi$, i.e.\    the adiabatic limit and the small field limit  are taken simultaneously. In \cite{MaT} we  prove Kubo's formula for the Hall conductivity for gapped magnetic Schr\"odinger Hamiltonians  using the NEASS approach. This requires to deal with the fact that the domain of $H_\epsi(t) = H_0 + \epsi f(\eta t) X_j$ changes at time $t=-1/\eta$ and to prove propagation estimates for time-dependent Stark-type Hamiltonians.

A breakthrough for interacting lattice systems was recently achieved by Bachmann, De Roeck, and Fraas  \cite{BDF}.
They prove the first adiabatic theorem for extended  lattice systems with local interactions that yields error estimates in local trace norms {\em uniformly in the size of the system}. This uniformity 
was the main mathematical challenge and the main innovation.
Their proof exploits locality  of interactions in the form of Lieb-Robinson propagation bounds \cite{LR} and the local inverse of the Liouvillian introduced by Hastings and Wen \cite{HW} (see also \cite{BMNS}). 
However, Bachmann et al.\  \cite{BDF} require the spectral gap not only for $H_0$, but also for the perturbed Hamiltonians $H_\epsi(t)$. In order to apply their result  to slowly varying potential perturbations that close the spectral gap (small fields over large regions), one needs to use the alternative gauge with a time-dependent vector potential and consider the adiabatic response instead, i.e.\ the first order deviation from ideal adiabatic behavior.  
This could be done using the results of \cite{MT}, which are a slight generalization of \cite{BDF}  in several directions: a super-adiabatic version of the theorem is formulated and proved which covers also the trace per unit-volume of extensive observables. This version is then used to derive Kubo's formula for conductance  and conductivity not from adiabatic switching of a small potential, but as the  adiabatic response for a Hamiltonian with time-dependent fluxes.

Finally, in \cite{Teu17} an adiabatic theorem for NEASSs in the setting of lattice systems with local interactions was established. It is shown that for perturbations by slowly varying potentials and/or by small local terms the above heuristics for NEASSs can be implemented rigorously by combining the techniques developed in \cite{BDF} with earlier ideas from space-adiabatic perturbation theory \cite{PST2}.
In the second part of this paper we show how to lift some results from \cite{Teu17} to infinite systems by taking a thermodynamic limit.

\subsection{Related results}\label{sectionC}

We end our review on linear response for gapped Hamiltonian systems  by briefly mentioning a few related results that did not easily fit into the previous categorization.

In a   series of works Bru and de Siqueira Pedra (see \cite{BD} and references therein) set up linear response theory and show how to evaluate Kubo's formula for interacting systems on the lattice in the thermodynamic limit. They do this on a very general level without taking an adiabatic limit and without any kind of gap assumption. Among many other results, 
 they show \eqref{Kubo1} with an error that is $o(\epsi)$ uniformly in the system size. The general tools they developed   for controlling the thermodynamic limit of interacting systems   turn out to be very useful  also for the problem discussed in the upcoming section, namely for the construction of NEASS in the thermodynamic limit. 

Linear response theory for (open) quantum systems from a  general perspective of quantum dynamical systems, discussing in particular also further consequences like Onsager relations, has been worked out by Jaksic, Pillet and collaborators in a series of works. Since we are not aware of a review paper on this topic, we just mention \cite{JOP}, which is probably closest to the setting of the present paper. 

There are clearly also physically relevant perturbations   that do not (or are not expected to) close the spectral gap of a gapped Hamiltonian, most notably, perturbations by small constant magnetic fields. The corresponding response coefficient is the magnetization.  As pointed out before, in this case the response coefficients can be determined from a family of equilibrium states $\rho_\epsi$  for the family of perturbed Hamiltonians $H_\epsi$, and one obtains
$
\sigma_A = \frac{\D}{\D \epsi} \langle  A\rangle_{\rho_\epsi}\big|_{\epsi = 0}
$
by taking a derivative of a family of equilibrium expectation values. 
For perturbations by constant magnetic fields proving existence of and  evaluating this derivative is still a highly non-trivial problem, since such perturbations are not within the realm of regular perturbation theory. Instead, a suitable magnetic perturbation theory was developed by Cornean and Nenciu \cite{CN} and applied, for example, in \cite{BCS} to derive and compute magnetic response coefficients in gapped non-interacting periodic systems. For mobility gapped systems on the lattice, magnetic response was considered also  in \cite{ST}. We are not aware of any adiabatic theorem applicable to the adiabatic switching of a time-dependent constant magnetic field, even in the gapped case.  

Let us finally mention Streda's formula from \cite{Streda}. Streda argued that in two dimensional systems the Hall conductivity 
equals the magnetic response for the particle density, i.e.\
$
\sigma_{12} =  \frac{\D}{\D B} \,\tau(\rho_B)  \big|_{B= 0}
$
in infinite non-interacting systems.
Bellissard \cite{Be88} showed that for discrete mobility gapped systems this definition gives the same value as the double-commutator formula, see also \cite{ST}. For 
unbounded Bloch-Landau Hamiltonians, Streda's formula was recently proved by Cornean, Monaco, and Moscolari \cite{CMM}, see also  \cite{CorneanStreda}. 
We are not aware of similar results on magnetic response for interacting systems.

 \section{Linear response for interacting fermions   in the thermodynamic limit}\label{section3}

In this section we  show how the results of \cite{Teu17} on the   justification of Kubo's formula using NEASSs for finite gapped systems at zero temperature can be lifted to infinite systems in the thermodynamic limit. It turns out that under suitable assumptions the NEASS $\Pi_\epsi$ discussed in the previous section exists also as a state on the algebra $\mathcal{A}_\Gamma$ of quasi-local observables for the infinite system and that $\Pi_\epsi$ is automorphically equivalent to the ground state $\rho_0$ of the unperturbed Hamiltonian $H_0$. Moreover, an adiabatic theorem that allows to formulate and prove Kubo's formula for the infinite system holds as well. To avoid technicalities, and because of limited space, we report here only the results for the special case of Hamiltonians of the form $H_\epsi(t) = H_0 + \epsi f(t) V$ and omit the proofs. The general statements and their proofs will be reported elsewhere~\cite{HT}. There we use results for controlling the thermodynamic limit that were worked out only quite recently in \cite{BD,NSY}.

\subsection{Fermions on the lattice: the mathematical framework}

We consider fermions with $s\in \N$   internal degrees of freedom (which could be the spin) on the lattice $\Gamma := \mathbb{Z}^d$. 
 Let $\mathcal{P}_0(\Gamma) = \set{X \subset \Gamma : \vert X\vert < \infty}$ denote the set of finite subsets of $\Gamma$. Then, for each $X \in \mathcal{P}_0(\Gamma)$, the corresponding 
 one-particle Hilbert space is $\mathfrak{h}_X := \ell^2(X, \C^s)$, the $N$-particle Hilbert space is its $N$-fold anti-symmetric tensor product   $\Hi_{X,N} := \bigwedge_{j=1}^N \mathfrak{h}_X$, 
and the fermionic Fock space is   $\mathfrak{F}_X := \bigoplus_{N=0}^{\dimfib |X|} \Hi_{X,N}$, 
where $\Hi_{X,0} := \C$. 
All these Hilbert spaces   are finite-dimensional and thus all linear operators on them  are bounded.
The local $C^*$-algebras $\mathcal{A}_X = \mathcal{L}(\mathfrak{F}_X)$ are generated by the identity element $\mathbf{1}_{X}$ and the  creation and annihilation operators $a_{x,i}^*, a_{x,i}$ for $x \in X$ and $1 \le i \le s$, which   satisfy the canonical anti-commutation relations (CAR), i.e.
 \[
\{a_{x,i},a_{y,j}\} = \{a_{x,i}^*,a_{y,j}^*\}=0, \quad \{a_{x,i},a_{y,j}^*\} = \delta_{x,y}\delta_{i,j}\mathbf{1}_X \quad \forall x,y \in X,\, 1\le i,j\le s\,.
\]
Here, $\{A,B\} = AB+BA$ denotes the anti-commutator of $A$ and $B$. If we have $X \subset X'$, $\mathcal{A}_{X}$ is naturally embedded as a sub-algebra of $\mathcal{A}_{X'}$. 
 We denote by $\mathcal{A}_{X}^N \subset \mathcal{A}_{X}$ the sub-algebra of elements commuting with the number operator $N_{X}= \sum_{x \in X} a_x^*a_x:= \sum_{x \in X} \sum_{i=1}^s a_{x,i}^*a_{x,i}$. As elements of $\mathcal{A}_{X}^N$ contain even numbers of creation and annihilation operators,  it holds that $[\mathcal{A}_{X}^N,\mathcal{A}_{X'}^N] = \{0\}$ whenever $X\cap X'=\emptyset$.
For the infinite system,   the local $C^*$-algebra is defined by the inductive limit 
\[
	\mathcal{A}_{\Gamma}^{\rm loc} := \bigcup_{X \in \mathcal{P}_0(\Gamma)} 	\mathcal{A}_{X}\qquad\mbox{and its closure is denoted by}\qquad \mathcal{A}_{\Gamma} := 	\overline{\mathcal{A}_{\Gamma}^{\rm loc}}^{\Vert \cdot \Vert}\,.
\]
In order to define families of operators that are sums of local terms, one uses the concept of ``interactions''. 
In the following we consider sequences of Hamiltonians defined on domains of the form
 $\Lambda(k) = \set{-k, ..., +k}^d$ with $k \in \mathbb{N}$.
A corresponding  interaction $\Phi = \set{\Phi^{\Lambda(k) }}_{k \in \mathbb{N}}$ for a fermionic system on the lattice $\Gamma$ is   defined as a family of maps
\begin{equation*}
\Phi^{\Lambda(k) } : \set{X \subset \Lambda(k)} \to  \mathcal{A}_{\Lambda(k)}^N\,,\quad   X \mapsto \Phi^{\Lambda(k) }(X) \in \mathcal{A}_{X}^N\subset  \mathcal{A}_{\Lambda(k)}^N
\end{equation*}
with values in the self-adjoint operators. 
The advantage of considering different maps $\Phi^{\Lambda(k) } $ for different $k \in \mathbb{N}$ instead of restrictions $\Phi|_{\Lambda(k)}$ of a single map $\Phi: \mathcal{P}_0(\Gamma)\to \mathcal{A}_\Gamma^{\rm loc}$ is the possibility to implement boundary conditions in order to model discrete tube or torus geometries. For example,  a   hopping term $a^*_{(k_1, \bar x)} a_{(-k_1, \bar x)}$   might only appear in the interaction for the  Hamiltonian  for that specific value of $k$ in order to connect opposite points on the boundary of  $\Lambda(k)$.
Moreover, we will also allow for different metrics $d^{\Lambda(k)}(\cdot, \cdot)$ on each $\Lambda(k)$ depending on  the intended geometry. For example, for a torus geometry opposite points on the boundary of $\Lambda(k)$ are considered neighbors and their distance is set to  one, while for a cube geometry their distance is $2k$.

The associated family of  self-adjoint operators $A= \{A^{\Lambda(k) }\}_{k\in \mathbb{N}_0}$ corresponding to an interaction $\Phi$ is defined by 
\begin{equation*}
A^{\Lambda(k) } = A^{\Lambda(k) }(\Phi) = \sum_{X \subset \Lambda(k)} \Phi^{\Lambda(k) } (X) \in \mathcal{A}_{\Lambda(k)}^N. 
\end{equation*}
We will  consider Hamiltonians $H_0$ that are operator-families given by  interactions $\Phi_{H_0}$ that are exponentially localized in the following sense: We say that an interaction is exponentially localized with rate $a>0$, if  
for all $n\in \N$ 
\begin{equation*}
\sup_{k\in \N}  \Vert \Phi^{\Lambda(k)}  \Vert_{a,n} := \sup_{k\in \N}  \sup_{x,y \in \Gamma} \sum_{\substack{ X \in \mathcal{P}_0(\Gamma): \\x,y\in X}} \mbox{$\Lambda(k)$-diam}(X)^n  \exp(a\cdot{d^{\Lambda(k)}(x,y)}) \, \Vert \Phi^{\Lambda(k) }(X)\Vert< \infty\,.
\end{equation*}
In this definition  we used implicitly that  for any interaction $\Phi$ the maps $\Phi^{\Lambda(k) }$ can   be extended to maps on all of  $\mathcal{P}_0(\Gamma)$ by declaring $\Phi^{\Lambda(k) }(X) = 0$, whenever $X \cap (\Gamma \setminus \Lambda(k)) \neq \emptyset$. This new mapping is called the extension of $\Phi^{\Lambda(k) }$ and is denoted by the same symbol. Similarly, given $\Phi^{\Lambda(k)}$ and $\Lambda(l) \subset \Lambda(k)$, we define the restriction $\Phi^{\Lambda(k) } |_{\Lambda(l)} : \set{X \subset \Lambda(l)} \to  \mathcal{A}_{\Lambda(l)}^N$ by
\begin{equation*}
	\Phi^{\Lambda(k) } |_{\Lambda(l)}(X) = \Phi^{\Lambda(k) }(X) \qquad \forall X \subset \Lambda(l). 
\end{equation*}

For the perturbation $V$ we will   consider     families of potentials $v = \set{v^{\Lambda(k)}: \Lambda(k) \to \mathbb{R}}_{k\in\N}$  that satisfy a uniform Lipschitz condition of the following type,
	\begin{equation*}
	C_v :=    \sup_{k \in \mathbb{N}} \sup_{x,y \in \Lambda(k)} \frac{\vert v^{\Lambda(k)}(x)- v^{\Lambda(k)}(y)\vert}{ d^{\Lambda(k)}(x,y)} < \infty\,,
	\end{equation*}
	and call them for short Lipschitz potentials.
With such a potential $v$ we associate the corresponding operator-family $V_v = \set{V_v^{  \Lambda(k)}}_{  k \in \mathbb{N}}$ defined by
$
	V_v^{ \Lambda(k)} = \sum_{x \in \Lambda(k)} v^{ \Lambda(k)}(x) a^*_xa_x
$.
Since in our definitions the functions $\Phi^{\Lambda(k)}$ resp.\ $v^{\Lambda(k)}$ defining an interaction resp.\ a potential can be, in principle, completely independent for different domains $\Lambda(k)$, we need to impose additional assumptions in order to guarantee the existence of a thermodynamic limit for all objects appearing in our construction. 
\begin{definition}
\balp
\item
An exponentially localized  interaction $\Phi = \set{\Phi^{\Lambda(k) }}_{k \in \mathbb{N}}$ is said to have a thermodynamic limit if it satisfies the following Cauchy-property:  
\begin{equation*}
\hspace{-1cm}\forall n\in \N\;\;\exists M_0 \in \N \;\; \forall M\ge M_0 \;\;\forall \delta >0 \;\; \exists K \ge M  \; \; \forall l , k \ge K: \,  \left\Vert \left( \Phi^{  \Lambda(l)}- \Phi^{  \Lambda(k)}\right) \big|_{\Lambda(M)}  \right\Vert_{a,n}  \le \delta\,. 
\end{equation*}
A family of operators is said to have a thermodynamic limit if and only if the corresponding interaction does. 
\item 
A    Lipschitz potential $v  = \set{v^{\Lambda(k) }}_{  k \in \mathbb{N}}$ is said to have a thermodynamic limit if  it is locally eventually independent of $k$, i.e.\ if
	\begin{equation*}
		\forall M\in \N  \;\; \exists K\ge M  \;\; \forall l , k \ge K: \, v^{  \Lambda(l)}  |_{\Lambda(M)} = v^{  \Lambda(k)}  |_{\Lambda(M)}. 
	\end{equation*}
	\ealp
\end{definition}
For a potential $v$ that has a thermodynamic limit,   the   point-wise limits 
$
v^{  \Lambda(k)}(x) \xrightarrow{k \to \infty} v^\infty (x)
$
exist for all $x\in \Gamma$ and
  the limiting function $v^\infty$ carries all important information about the potential as far as the thermodynamic limit is concerned.

A relevant example for a Hamiltonian $H_0$ that is exponentially localized and has a thermodynamic limit   is
\begin{equation}\label{example:HTPHW}
H^\Lambda_{0}   =  \sum_{ (x,y)\in \Lambda^2}  \hspace{-8pt}  a^*_x \,T( x\lminus y) \,a_y+ \sum_{x\in\Lambda}  a^*_x\phi( x)a_x  
  + \hspace{-4pt} \sum_{\{x,y\}\subset \Lambda } \hspace{-8pt} a^*_xa_x \,W( d^\Lambda(x,y))\,a^*_ya_y - \mu \,N_\Lambda\,.
\end{equation}
Here we assume that the kinetic term $T :\Gamma \to \mathcal{L}(\C^\dimfib)$ is an exponentially fast decaying   function with $T( -x) = T(  x)^*$, the potential term $\phi :\Gamma \to \mathcal{L}(\C^\dimfib)$ is a bounded function taking values in the self-adjoint matrices,  and  the two-body interaction $W :[0,\infty)\to 
\mathcal{L}(\C^\dimfib)$ is exponentially decaying  and also takes values in the self-adjoint matrices. 
Note that $ x\lminus y$ in the kinetic term in \eqref{example:HTPHW} refers to the difference modulo $\Lambda(k)$ if $\Lambda(k)$ is supposed to have a torus geometry.

The most relevant   Lipschitz potentials we have in mind are the linear potential $v^{\Lambda(k)}_D(x) := x_i $ for some $i\in\{1,\ldots, d\}$ if the metrics $d^{\Lambda(k)}$ correspond to a cube geometry (think of Dirichlet boundary conditions on the box) or the saw-tooth potential 
\[
v^{\Lambda(k)}_P(x) := \left\{ \begin{array}{cl}  
x_i & \mbox{if $x_i\in [-\frac{k}{2},\frac{k}{2}]$}\\
k-x_i& \mbox{if $x_i\in (\frac{k}{2}, k]$}
\\
-k-x_i& \mbox{if $x_i\in [-k,- \frac{k}{2})$}
\end{array}\right.
\]
for the torus geometry. Note that $C_{v_D}= C_{v_P} = 1$ and that  both potentials have a thermodynamic limit and converge point-wise to the same function $v^\infty(x) =x_i$. As we will see, they also define the same response coefficients in the thermodynamic limit when added to a gapped Hamiltonian of the type \eqref{example:HTPHW}.

The following proposition can be proved exactly as Theorem~3.5  in \cite{NSY}  using also Theorem~3.4 and Theorem~3.8 from the same reference.  It shows, that the   property of having a thermodynamic limit for the interaction resp.\ the potential  guarantees also the existence of a thermodynamic limit for the associated evolution operators. 

\begin{proposition}{\bf Thermodynamic limit of Cauchy-interactions \cite{NSY}}\\
Let $H_0$ and $H_1$ be operator-families that are associated with two exponentially localized interactions     and let $v$ be a Lipschitz potential with associated operator-family~$V_v$, all having a thermodynamic limit. Let $V:= V_v+H_1$.
\balp
\item Let $H_\epsi := H_0 + \epsi V$ for some $\epsi\in \R$. Then there exists a unique one-parameter group $t\mapsto \E^{\I\mathcal{L}_{H_\epsi} t}: \mathcal{A}_\Gamma\to \mathcal{A}_\Gamma$ of automorphisms such that for all $A\in \mathcal{A}_\Gamma^{\rm loc}$
\[
 \E^{\I\mathcal{L}_{H_\epsi} t}[A] = \lim_{k\to \infty} \E^{\I H_\epsi^{\Lambda(k)} t} \,A\, \E^{-\I H_\epsi^{\Lambda(k)} t}
\]
\item Let $f:\R\to \R$ be  smooth  and put $H_{\epsi,\eta} (t) := H_0 + \epsi f(\eta t) V$ for some $\epsi\in \R$. 
Denote by $U^{\Lambda(k),\eta}(t,t_0)$ the  evolution family generated by $H_{\epsi,\eta} (t)$, i.e.\ the solution to the Schr\"odinger equation
\[
\I \tfrac{\D}{\D t} U^{\Lambda(k),\eta}(t,t_0) = H_{\epsi,\eta}^{\Lambda(k)} (t)\,U^{\Lambda(k),\eta}(t,t_0)  \qquad\mbox{with } \quad U^{\Lambda(k),\eta}(t_0,t_0) = {\rm Id}\,.
\]
Then there exists a unique co-cycle of automorphisms $\mathfrak{U}^{\eta}_{t,t_0}:\mathcal{A}_\Gamma\to \mathcal{A}_\Gamma$ such that for all $A\in \mathcal{A}_\Gamma^{\rm loc}$
\[
\mathfrak{U}^{\eta}_{t,t_0}[A] = \lim_{k\to \infty} U^{\Lambda(k),\eta}(t,t_0)^*   \,A\, U^{\Lambda(k),\eta}(t,t_0)  \,.
\]
\ealp
\end{proposition}
The main additional assumption on the Hamiltonian $H_0$ that we need is the {\em gap assumption}: We say that the operator-family $H_0=\{ H_0^{\Lambda(k)}\}_{k\in\N}$ has a simple gapped ground state, if there exists  $L \in \mathbb{N}$ such that for all $k \ge L$ and corresponding $\Lambda(k)$ the smallest eigenvalue $E_0^{\Lambda(k)}$ of the operator $H_0^{\Lambda(k)}$ is simple and the spectral gap is uniform in the system size, i.e. there exists   $g>0$ such  dist$(E_0^{\Lambda(k)}, \sigma(H_0^{\Lambda(k)})\setminus \{E_0^{\Lambda(k)}\}) \ge g >0 $ for all $k \ge L$. 

In addition, we require that also the sequence of ground states $\rho_0^{\Lambda(k)}$ has a thermodynamic limit. To formulate this condition, recall that
a normalized positive linear functional $\omega$ on the $C^*$-algebra $\mathcal{A}_\Gamma$   is called a state.  
Here normalized means $\omega({\rm id})=1$ and positive means
$
\omega(A^*A) \ge 0
$
for all $A \in \mathcal{A}_\Gamma$. 
By the Banach-Alaoglu theorem,  the set of states $E_{\mathcal{A}_\Gamma}$ is weak$^*$-compact.
If $\mathcal{B}$ is a subalgebra of $\mathcal{A}_\Gamma$ and $\omega$ is a state on $\mathcal{B}$, then there exists a state $\hat{\omega}$ on $\mathcal{A}_\Gamma$ which extends $\omega$ by the Hahn-Banach theorem. 
In this way, the states $B\mapsto \tr(\rho_0^{\Lambda(k)} B)$ can be extended to the whole algebra $\mathcal{A}_{\Gamma}$ and are denoted by the same symbol $\rho_0^{\Lambda(k)}$.
To avoid the extraction of subsequences,  we will assume that the sequence of ground states  $(\rho_0^{\Lambda(k)})_{k\in\N}$ converges to a unique limiting point $\rho_0 \in E_{\mathcal{A}_\Gamma}$.
 \\[0mm]

\noindent
\textbf{(A1) Assumptions on $H_0$.}
\\{\em
Let $H_0$ be the Hamiltonian of an exponentially localized interaction that has a thermodynamic limit. We assume that $H_0$ has a gapped ground state such that the sequence of ground states $(\rho_0^{\Lambda(k)})_{k \in \mathbb{N}_0}$  converges in the weak$^*$-topology  to a state  
$\rho_0$ on~$\mathcal{A}_\Gamma$.}\\[0mm]

\noindent
\textbf{(A2) Assumptions on the perturbation.}
\\{\em 
Let $V= V_v + H_1$, where
$v$ is a Lipschitz potential  and  $V_v$ denotes the corresponding operator-family, and 
 $H_1$ is the Hamiltonian of an exponentially localized interaction, both having   a thermodynamic limit. }
\\

Note that for  non-interacting Hamiltonians $H_0$ of the type \eqref{example:HTPHW} on a torus, i.e.\  with $W\equiv0$,
condition (A1) is satisfied whenever the chemical potential $\mu$ lies in a gap of the spectrum of the corresponding one-body Hamiltonian operator on the infinite domain, a condition that can be checked easily. 
And  it was recently shown in \cite{H,DS}, that for sufficiently small  $W\not=0$ the spectral gap remains open. 

We have now all prerequisites to formulate our main results on the existence and properties of NEASS for interacting systems in the thermodynamic limit. They are all adaptions of corresponding results for finite systems proved in \cite{Teu17}. As they are not simple corollaries of \cite{Teu17} but require some careful adaptions, their proofs will be given elsewhere \cite{HT}.

The first theorem states that under Assumptions (A1) and (A2) there exists a NEASS for the perturbed Hamiltonian $H_\epsi = H_0 + \epsi V$   close to the 
ground state  of~$H_0$ (cf.\ Theorem~3.1 in \cite{Teu17}).

\begin{theorem}\label{SpaceAdiabaticThm}{\bf Existence of NEASSs}\\
 Let the Hamiltonian $H_\epsi  =H_0 + \epsi V$ satisfy (A1) and (A2).  Then for any  $\epsi\in[-1,1]$ there exists a near-identity  automorphism $\alpha_\epsi$ of $\mathcal{A}_\Gamma$  such that the state $\Pi_\epsi$ defined by
 \[
 \Pi_\epsi(A)  := \rho_0(\alpha_\epsi[A]) \qquad \mbox{for all } A\in\mathcal{A}_\Gamma
 \]
 is almost-invariant in the following sense:
 for any $n\in \N$ there exists a constant  $C_n >0$ such that for all  finite $X\subset \Gamma$, $A\in  \mathcal{A}_X^N\subset \mathcal{A}_\Gamma$, and $\epsi\in [-1,1]$  
  \begin{equation}\label{AdiState00}
 \left|  \Pi_\epsi( \E^{\I \mathcal{L}_{H_\epsi }t}[A] )-  \Pi_\epsi(A) \right| \leq C_n \, {|\epsi|^{n  }}    (1+|t|^{2d}) \,|X|^3\,\|A\|
\,.
\end{equation}
By near-identity we mean that $\alpha_\epsi$ is of the form $\alpha_\epsi = \E^{ \I \epsi \mathcal{L}_{S_\epsi}}$ for an almost exponentially localized operator family $S_\epsi$ that has a thermodynamic limit.
 \end{theorem}
 
The next theorem is a special case of a more general adiabatic theorem for NEASS, cf.\ Theorem~5.1 in \cite{Teu17}. It shows that when adiabatically switching on the perturbation, then the initial ground state of $H_0$ dynamically evolves up to small errors into the corresponding NEASS $\Pi_\epsi$ for the perturbed Hamiltonian $H_\epsi$ as long as the adiabatic parameter $\eta$ is small but not too small (see also Proposition 3.2 in \cite{Teu17}).

 \begin{theorem}\label{prop:switch}{\bf Adiabatic switching}\\
 Let the Hamiltonian $H_\epsi  =H_0 +  \epsi V$ satisfy (A1) and (A2).   Let   $f:\R \to \R$ be a smooth ``switching'' function with   
  $f(t) =0$ for $t\leq -1$ and $f(t) = 1$ for $t\geq  0$, and define  $H_{\epsi,\eta}(t) := H_0 + \epsi f(\eta t) V$. 
  Let $\mathfrak{U}^{ \eta}_{t,t_0}$ be the Heisenberg time-evolution on $\mathcal{A}_\Gamma$ generated by $H_{\epsi,\eta}(t)$ with adiabatic parameter $\eta\in (0,1]$.

Then
  for any $n>d$ there exists a constant $C_n$ such that for any  finite $X\subset \Gamma$ and $A\in  \mathcal{A}_X^N\subset \mathcal{A}_\Gamma$   and  for all $t\geq 0$  
  \begin{equation} \label{NEASSerror}
 \left| 
 \rho_0( \mathfrak{U}^{ \eta}_{t/\eta,-1/\eta}[A]) - \Pi_\epsi(A) 
\right|   \leq \;\frac{|\epsi|^{n+1} + \eta^{n+1}}{\eta^{d+1}}  \,C_n \,(1+t^{d+1})  \,|X|^2\,\|A\|
\,,
\end{equation}
where $\Pi_{\epsi } $ is the NEASS of $H_{\epsi }$ constructed in Theorem~\ref{SpaceAdiabaticThm}.
 \end{theorem}
Note that 
\eqref{NEASSerror} shows that, as long as the adiabatic parameter $\eta$ satisfies 
$
1\gg \eta \gg |\epsi|^\frac{n+1}{d+1}  
$
for some $n>d$,
the initial ground state $\rho_0$ of $H_0$ evolves, up to a small error, into the NEASS $\Pi_{\epsi} $ that is independent of the form of the switching function $f$ and of~$\eta$. 
Slower switching must be excluded, because, in general, the NEASS is an almost-invariant but not an invariant state for the instantaneous Hamiltonian. Its life-time depends on the strength of the perturbation, i.e.\ on $\epsi$, and it is thus  not surprising that the relevant time scale for the adiabatic switching process depends on $\epsi$ as well. 
 
In order to compute response coefficients, we need to expand $\Pi_\epsi$ in powers of $\epsi$ (cf.\ Proposition 3.1 in \cite{Teu17}).

\begin{proposition}\label{proposition:expand1}{\bf Asymptotic expansion of the NEASS}\\
Under the assumptions of Theorem~\ref{SpaceAdiabaticThm}
there exist linear maps $\mathcal{K}_j: \mathcal{A}_\Gamma^{\rm loc} \to  \mathcal{A}_\Gamma$, $j\in\N$, such that for any $n\in\N $   there is a constant $C_n$ such that
for any    finite $X\subset \Gamma$ and $A\in  \mathcal{A}_X^N\subset \mathcal{A}_\Gamma$ it holds that  
\[
 \left|   \Pi_\epsi(A)  -   \sum_{j=0}^n \epsi^j\,
\rho_0(  \mathcal{K}_j [A] )
\right| \;\leq\; C_n\,|\epsi|^{n+1}\,|X|^2\,\|A\| \,.
\]
The constant term is $\mathcal{K}_0=\mathbf{1}_{\mathcal{A}_\Gamma}$, which shows that $\alpha_\epsi$ is indeed a near-identity automorphism when $|\epsi|\ll 1$. The linear term $\mathcal{K}_1$ is a densely defined derivation on $\mathcal{A}_\Gamma$ that satisfies
\begin{eqnarray}\label{interactingKubo}
\rho_0(  \mathcal{K}_1  [A] ) &=& -   \, \lim_{k\to \infty}    \left\langle \left[  \mathcal{I}_{H_0^{\Lambda(k)}} (V^{\Lambda(k)}) , 
A \right]   \right\rangle_{\rho_0^{\Lambda(k)} }\nonumber\\&=&
  \lim_{k\to \infty} \tr \left(  \mathcal{L}_{H_0^{\Lambda(k)}}^{-1}( [V^{\Lambda(k)},\rho_0^{\Lambda(k)}])\,
A\right)
 \end{eqnarray}
for every $A\in \mathcal{A}_\Gamma^{\rm loc}$. Here $ \mathcal{I}_{H_0^{\Lambda(k)}} $ is a local version of the inverse Liouvillian introduced in \cite{HW} and from the first expression in \eqref{interactingKubo} it follows that $\rho_0(  \mathcal{K}_1  [A] ) $ depends on $V_v$ only through the limiting function $v^\infty$.
 \end{proposition}

Finally, we can combine the previous results in order to formulate our main theorem about linear and higher order response for gapped interacting systems in the thermodynamic limit (cf.\ Theorem~4.1 in \cite{Teu17}).

\begin{theorem}\label{response}{\bf Linear and higher order response}\\
Under the same assumptions as in Theorem~\ref{prop:switch},
let again $\mathfrak{U}^{ \eta}_{t,t_0}$ be the Heisenberg time-evolution on $\mathcal{A}_\Gamma$ generated by $H_{\epsi,\eta}(t)$ with adiabatic parameter $\eta\in (0,1]$.
 For   $A\in \mathcal{A}_\Gamma$   define   the  total response  as
 \[
 \Sigma^{\epsi,\eta,f}_A(t) :=  \rho_0( \mathfrak{U}^{\eta}_{t/\eta,-1/\eta}[A])   -
 \rho_0(A) 
 \,,
 \]
 and for $j\in\N$ the $j$th order response coefficient as
 $
 \sigma_{A,j} := 
\rho_0( \mathcal{K}_j [A ])
 $,
 where the $\mathcal{K}_j $'s were defined   in Proposition~\ref{proposition:expand1} and 
 $\sigma_{A,1}$ is explicitly given by  \eqref{interactingKubo}, i.e.\ by  the thermodynamic limit of Kubo's formula \eqref{Kubo3}.
 
 Then for any $n,m \in\N$    there exists a constant $C_{n,m} $ independent of   $\epsi$, such that for  any    finite $X\subset \Gamma$ and $A\in  \mathcal{A}_X^N\subset \mathcal{A}_\Gamma$ and all $t\geq 0$
 \begin{equation}\label{eq:response}
  \sup_{\eta\,\in\left[|\epsi|^m,\,|\epsi|^\frac{1}{m}\right]} \left|  \Sigma^{\epsi, \eta,f}_A(t) 
 -   \sum_{j=1}^n \epsi^j  \sigma _{A,j}
 \right| \leq |\epsi|^{n+1}\,C_{n,m}\,(1+t^{d+1})\,|X|^2\,\|A\| \,.
 \end{equation}
\end{theorem}  
Note that the condition that $\eta \in [|\epsi|^m,|\epsi|^\frac{1}{m} ]$ for some $m\in\N$ makes sure that the switching is neither too slow ($\eta\geq |\epsi|^m$) nor too fast 
($\eta\leq |\epsi|^\frac{1}{m}$). Too slow switching  would be switching on time-scales longer than the life-time of the NEASS, while too fast switching would no longer allow for an expansion of the total response in powers of $\epsi$.

While we believe that this result and the NEASS approach in general are an important contribution to the mathematical understanding of linear response theory for transport in gapped Hamiltonian systems, there are still several open questions: 
An obvious conjecture would be that our results remain valid if one replaces   the gap condition for the local Hamiltonians in Assumption (A1) by a gap condition for the Hamiltonian for the infinite system (in the GNS representation). At least for the case that $V\in  \mathcal{A}_\Gamma$, we expect that the methods recently developed in \cite{MO} can be used to adapt our proofs accordingly. 

A presumably much harder   but physically more interesting problem is to justify Kubo's formula for current response in situations where  $H_0$ no longer has a spectral gap but only a mobility gap. Even for non-interacting systems we know of no results in this direction yet.

\end{document}